\documentclass[aps,prd,longbibliography ,twocolumn,nofootinbib,10pt]{revtex4-2}
\usepackage{graphicx, comment}
\usepackage{amsmath}
\usepackage{amsfonts}
\usepackage{graphicx}
\usepackage{dsfont}
\usepackage[german, english]{babel}
\usepackage{amssymb}
\usepackage{ifthen}
\usepackage{ulem}
\usepackage{color}
\usepackage{float}
\usepackage{physics }
\usepackage{hyperref}
\usepackage{bbold}
\usepackage{lipsum}
\usepackage{nccmath}
\usepackage{xcolor}

\newcommand{\br}{\boldsymbol{r}}
\newcommand{\be}{\boldsymbol{e}}

\newcommand{\bs}{\boldsymbol{s}}
\newcommand{\bt}{\boldsymbol{t}}

\newcommand{\bp}{\boldsymbol{p}}
\newcommand{\bP}{\boldsymbol{P}}

\newcommand{\bX}{\boldsymbol{X}}
\newcommand{\bzero}{\boldsymbol{0}}
\newcommand{\bmu}{\boldsymbol{\mu}}
\newcommand{\mycomment}[1]{}

\begin{document}
\begin{abstract}
We study ferroelectric domain walls in barium titanate.
We search for structurally nontrivial, so-called non-Ising domain walls, where the Polarisation is non-zero along the entire wall. 
Our approach enables us to find solutions for domain walls in any orientation, and the existence and energy of these walls depend on their particular orientation.
We find that, across all phases of the material, there are orientations where the non-Ising walls have lower energy than Ising walls.
The most interesting property of these domain walls is their non-monotonic interaction forces, allowing them to form stable domain-wall clusters rather than following standard behavior where domain walls annihilate or repel each other.
We found the required external electric field to create the non-Ising configurations.
Besides theoretical interest, this unconventional property of domain walls makes them a good candidate for memory application.
\end{abstract}
\title{Ferroelectric domain wall clusters in barium titanate}
\author{Chris Halcrow}
\email{chalcrow@kth.se}
\author{Egor Babaev}
\affiliation{Department of Physics, KTH-Royal Institute of Technology, SE-10691, Stockholm, Sweden}
\date{\today}

\maketitle

\section{Introduction}

Domain walls (DWs) are ubiquitous in ferroelectric materials.
In an Ising wall, the most common type, the electric polarisation vanishes at the center of the wall.
It was once thought that all DWs in barium titanate were Ising, but more recent theory has predicted the existence of exotic walls in different phases of the material. There are high-energy``Bloch" wall in the tetragonal phase\cite{huang1997} and low-energy non-Ising walls in the  rhombohedral phase \cite{taherinejad2012}.
These DWs have a non-zero polarisation vector at all points in space, making them analogous to Bloch and Néel walls in magnetic systems. 
Novel ``non-Ising" DWs have been theoretically predicted  \cite{lee2009, gu20147} and reported experimentally \cite{wei2016Neel, katia2017non}. 
The energy of the walls depend on their orientation, showing the importance of geometric considerations.

The diversity, robustness, mobility and sharpness of their domain walls make ferroelectrics candidates for promising new technological applications. 
One significant application is in memory, where the presence or absence of the DW can represent the binary state. 
The creation and non-destructive readability of DW states has been demonstrated in various systems \cite{sharma2017nonvolatile, sun2022nonvolatile, ma2018controllable, jiang2018temporary}. 
Hence ferroelectrics represent not only an academically interesting platform to search for new kinds of topological defects, but also have great potential for new applications. 
However, DW-racetrack memory has a major drawback compared to other technologies: pairs of domain walls are generically unstable to annihilation. 
Usually, it is assumed that the DWs are pinned to a defect or kept apart to avoid this instability.

In this paper, we report the existence of stable wall-wall pairs and clusters in barium-titanate. In soliton language, these configurations are clusters of soliton-antisoliton pairs. 
Stable configurations of this type with net zero topological charge are incredibly rare in condensed matter. Some examples are: skyrmionium and skyrme bags, made from a placing skyrmions inside an anti-skyrmion, have been predicted and observed in magnetic systems \cite{finazzi2013laser, zhang2016control, zhang2018real, rybakov2019chiral, foster2019two} and stable vortex-antivortex excitations have been generated numerically in a spin imbalanced superfluid \cite{barkman2020ring} and noncentrosymmetric
superconductors  \cite{garaud2020vortices,samoilenka2020spiral}.
Some higher-dimension dimensional solitons can be interpreted as composite lower-dimensional solion-antisolitons, such as the the skyrmions in the form of co-centric domain and anti-domain-wall rings decorated by vortices in certain superconductors \cite{garaud2013chiral}. In that example the decoration of domain walls by vortices enforces the stability.
We are not aware of a condensed matter example of systems allowing energetically stable domain-antidomain-wall structures.
In ferroelectrics there are examples of stable wall-wall pairs in periodic systems \cite{taherinejad2012bloch, hu1997computer, hu1998three}, but these boundary conditions stabilise pairs which repel each other. Ours attract at long range and are thus true local minima.
The stability of the domain wall clusters that we find crucially relies on the wall's non-Ising character and the geometry of the system. 

To study these configurations, we first generalise the Ginzburg-Landau-Devonshire formalism of \cite{MRH2010} to consider walls with arbitrary orientations.
We demonstrate that there are wall-orientations which support non-Ising DWs in every phase of barium-titanate.
We also demonstrate how the stable configurations can be created using an external electric field. During our analysis, we apply adapt several tools from soliton theory to the field of ferroelectrics.
These include Arrested Newton flow, the string method and a small-fluctuation gradient flow.
We aim to explain these methods in a clear and general manner, so that they can be readily applied to problems in higher dimensions.

\section{Theoretical framework}

The Ginzburg-Landau-Devonshrie model of ferroelectrics features the Polarisation vector $\bP = (P_1, P_2, P_3)$ and the symmetric $3\times 3$ strain tensor $u_{ij}$ as order parameters. Throughout derivatives are denoted by $\tfrac{\partial}{\partial x_a} = \partial_a$ and we sum over repeated indices. We study a free energy which is compatible with the symmetries of the underlying crystal. In our case, the goal is to minimise the free energy given by
\begin{align*}
	F = &\tfrac{1}{2}G^o_{abcd}\partial_a P_b \partial_c P_d + V^o(P) \\
	&+ \tfrac{1}{2}C^o_{abcd}u_{ab}u_{cd} - q^o_{abcd}u_{ab}P_cP_d \\
	V^o = &A^o_{ab}P_aP_b + A^o_{abcd}P_aP_bP_cP_d + A^o_{abcdef}P_aP_bP_cP_dP_eP_f
\end{align*}
subject to the elastic compatibility condition
\begin{equation} \label{eq:compat}
	\mathcal{D}u = \epsilon_{acd}\epsilon_{bef}\partial_{c}\partial_e u_{df} = 0 \, .
\end{equation}
We use the superscript ``o" to mean ``original coordinates". 

We will sometimes use Voigt notation for the strain tensor. Here, we reshape the symmetric matrix $u$ into a 6-vector
\begin{eqnarray} \label{eq:Voigtep}
	( \epsilon_1, \epsilon_2, \epsilon_3, \epsilon_4, \epsilon_5, \epsilon_6 ) = ( u_{11}, u_{22}, u_{33}, 2u_{23},2u_{13}, 2u_{12} )\, .
\end{eqnarray}
Latin and Greek letters are used when summing over the old and new subscripts, so that $a=1,2,3$ but $\alpha=1,...,6$. Provided the tensors are symmetrised, the Voigt coefficients are uniquely defined by
\begin{align}
	C^o_{\alpha \beta}\epsilon_\alpha \epsilon_\beta &= C^o_{abcd}u_{ab}u_{cd} \\
	q^o_{\alpha cd} \epsilon_{\alpha} &= q^o_{abcd}u_{ab} \, .
\end{align}
Using this notation, we can immediately complete the square on the part of $F$ that depends on strain, which becomes
\begin{align} \label{eq:completethesquare}
	\tfrac{1}{2}C^o_{\alpha\beta}&\left( \epsilon_\alpha - Q^o_{\alpha ef}P_eP_f\right)\left( \epsilon_\beta - Q^o_{\beta gh}P_gP_h\right) \nonumber \\
	&- \tfrac{1}{2}C^o_{\alpha \beta}Q^o_{\alpha ef}Q^o_{\beta gh}P_eP_fP_gP_h \, .
\end{align}
The tensor $Q$ is given by 
\begin{equation}
 Q^o_{\alpha b c} = (C^o)^{-1}_{\alpha \beta}q^o_{\beta b c}
\end{equation}
Experimental works determine $Q$ directly, rather than $q$. We can absorb the leftover part of \eqref{eq:completethesquare} into $V^o$, giving the effective potential
\begin{equation}
	\tilde{V}^o(P) = V^o(P) - \tfrac{1}{2}C^o_{\alpha \beta}Q^o_{\alpha ef}Q^o_{\beta gh}P_eP_fP_gP_h \, .
\end{equation}
Note that the effective potential $\tilde{V}^o$ only differs from $V^o$ in its quartic dependence on $P$.

First, consider the global minima of the free energy $(P^V,\epsilon^V)$. 
In what follows we will refer to these homogeneous states as {\it vacua}.
The system has multiple vacua, which satisfy
\begin{align} \label{eq:vaceq}
	\frac{\partial \tilde{V}^o}{\partial P_a}\bigg\rvert_{P=P^V} = 0\, , \qquad 
	\epsilon^V_{\alpha} = Q^o_{\alpha cd}P^V_bP^V_c \, .
\end{align}
In this paper, we focus on domain walls: configurations which connect two vacua in physical space and only vary in one direction. To take advantage of this, we'll now change coordinates to $(\bs, \bt, \br)$, with $\bs$ in the direction of the wall-normal. The rotation matrix is given by
\begin{equation}
	R = \left( \bs, \bt, \br \right)^T
\end{equation}
We take the important coordinate, $\bs$ the wall-normal, as the {\it first} basis vector in our new coordinates. Since we are using tensor notation, the material constants transform in a simple way. For example,
\begin{equation}
	A^o_{abcd} \to A_{abcd} = R_{ae}R_{bf}R_{cg}R_{dh}A^o_{efgh} \, ,
\end{equation}
This transformation also applies to $C_{abcd}$ but not $C_{\alpha\beta}$. As such, we apply the rotation in the `$abcd$'-coordinates and then transform to Voigt notation using \eqref{eq:Voigtep}. We denote the material constants in these new coordinates without any superscript. Note that the material constants depend on the orientation of the wall. 

Since the fields only vary in the $\bs$ direction the compatibility condition \eqref{eq:compat} simplifies significantly: $\epsilon_2$, $\epsilon_3$ and $\epsilon_4$ must be constant along the wall ($\epsilon_1$ is special, as the wall-normal $\bs$ is in the `1' direction in our new coordinate). This is important. If the Polarisation vector changes along the wall, the elastic vacuum condition \eqref{eq:vaceq} can fail and the effective theory for the Polarisation vector is modified, sometimes significantly.

We'll consider infinite domain walls, which should have finite energy. Hence the vacuum conditions \eqref{eq:vaceq} must be satisfied at both ends of the wall.

Suppose the domain wall connects two vacua   $P^{V_{-\infty}}$ and $P^{V_\infty}$. The strain is fixed at $-\infty$ as
\begin{equation} \label{eq:strainminus}
	\epsilon_\alpha^{V_{-\infty}} = Q_{\alpha b c} P^{V_{-\infty}}_b P^{V_{-\infty}}_c \, .
\end{equation}
Now the tensors have no ``o" superscript as we have rotated into the new basis. Similarly, the strain at $+\infty$ is given by
\begin{equation} \label{eq:strainplus}
	\epsilon^{V_\infty}_\alpha = Q_{\alpha b c} P^{V_{\infty}}_b P^{V_{\infty}}_c \, .
\end{equation}
But the elastic compatibility condition means that the strains perpendicular to the wall are constant, so that $\epsilon_{2/3/4} = \epsilon^V_{2/3/4}$. In particular, these strains at either end of the wall must be equal:
\begin{equation} \label{eq:constcond}
	Q_{\alpha b c} P^{V_{-\infty}}_b P^{V_{-\infty}}_c = Q_{\alpha b c} P^{V_{\infty}}_b P^{V_{\infty}}_c \, , \quad \alpha = 2,3,4 \, .
\end{equation}
Note that $Q$ is a function of wall orientation. We'll think of this equation as follows: given ingoing and outgoing vacua, which wall orientations satisfy \eqref{eq:constcond} and hence give a permissible wall? We will see an example in the next section.

Now, given the in and out Polarizations and a permissible wall, we would like to minimise the free energy. This would be simple if we could set $\epsilon(s)_\alpha= Q_{\alpha bc} P_b(s)P_c(s)$. Then the potential energy would  be $\tilde{V}$. However, the elastic compatibility condition means that some of the $\epsilon$ must be constant, so cannot depend on $P$ in this way. Instead, we first minimise for the constant strains $\epsilon_{2/3/4}$ . To help, we split $C$ and $q$ into parts parallel and perpendicular to the wall
\begin{align}
	q_{\alpha b c} &= q_{\alpha b c}^\parallel + q_{\alpha b c}^\perp \\
	C_{\alpha\beta} &= C^\parallel_{\alpha\beta} + C_{\alpha\beta}^\perp + C^m_{\alpha\beta} 
	\end{align}
where $q^\perp_{\alpha bc}$ is only non-zero when $\alpha=2,3,4$, $q^\parallel_{\alpha b c}$ is non-zero when $\alpha=1,5,6$, $C^\perp_{\alpha\beta}$ is non-zero when $\alpha$ and $\beta$ both belong to $\{2,3,4\}$; $C^{||}$ is only non-zero when both belong to $\{1,5,6\}$; and $C^m$ is non-zero otherwise. As seen earlier, the perpendicular strains satisfy
\begin{equation}
	\epsilon_\alpha = \epsilon^V_\alpha = Q_{\alpha b c}P_b^{V_{\infty}}P_c^{V_{\infty}} \, ,\quad \alpha = 2,3,4 \, .
\end{equation}
The parallel strains are free to minimise the part of the energy which depends on them,
\begin{equation}
	\tfrac{1}{2}C^\parallel_{\alpha\beta} \epsilon_\alpha \epsilon_\beta + C^m_{\alpha\beta}\epsilon_\alpha \epsilon^V_\beta - q^\parallel_{\alpha b c} \epsilon_\alpha P_b P_c\,  ,
\end{equation}
which is minimal when
\begin{equation} \label{eq:nosense}
	\epsilon_\alpha = \left(C^{||}\right)_{\alpha \beta}^{-1}\left( q^{||}_{\beta b c}P_bP_c - C^m_{\beta \gamma} \epsilon^V_\gamma \right) \, .
\end{equation}
Here the inverse of $C^{||}$ is defined as the inverse of the $3\times 3$ non-zero part, projected back into its $6 \times 6$ form. This is the unique matrix which satisfies 
\begin{equation}
    \left( C^{||}\left(C^{||}\right)^{-1}\right)_{\alpha \beta} = \,\begin{cases} \delta_{\alpha \beta} \quad \alpha = 1,5,6\\
    0 \quad \text{otherwise}\end{cases}
\end{equation} 
We can make more sense of equation \eqref{eq:nosense} using the fact that $q^{||} = C^{||}Q^{||} + C^mQ^{\perp}$. Then
\begin{equation}
	\epsilon_\alpha = Q^{||}_{\alpha b c} P_bP_c + \left(C^\parallel\right)^{-1}_{\alpha\beta}C^m_{\beta\gamma}\left(Q^\perp_{\gamma b c}P_bP_c  -\epsilon^V_\gamma  \right)
\end{equation}
for $\alpha=1,5,6$. As $s \to \pm \infty$, the second term vanishes and these strains also approach their vacuum values \eqref{eq:vaceq}, as we would expect.

Overall, the minimal energy strains are given by
\begin{equation} \label{eq:uHK}
	\epsilon_\alpha = H_{\alpha bc}P_bP_c + K_{\alpha} 
\end{equation}
where
\begin{align}
	H_{\alpha bc} &=  \begin{cases} Q^{||}_{\alpha b c} + \left(C^\parallel\right)^{-1}_{\alpha\beta}C^m_{\beta\gamma} Q^\perp_{\gamma b c}\quad &\alpha=1,5,6\\
		0 \quad &\alpha =2,3,4\end{cases} \\
	K_\alpha &=\begin{cases}
		-\left(C^{\parallel}\right)^{-1}_{\alpha\beta}C^{m}_{\beta\gamma}\epsilon^V_\gamma \quad &\alpha=1,5,6\\
		\epsilon^V_\alpha \quad &\alpha = 2,3,4 \, .
	\end{cases} 
\end{align}
Finally, we substitute these expressions into the free energy to find an effective free energy for $P$, which depends on a single coordinate $s$. It is
\begin{align} \label{eq:finalF}
	F = \,\,&\tfrac{1}{2}G_{sasb}\partial_s P_a\partial_s P_b + V(P)\\
	V = \,\,&\tilde{A}_0 + \tilde{A}_{ab}P_a P_b + \tilde{A}_{abcd}P_a P_b \nonumber \\&+ A_{abcdef}P_a P_b P_c P_d P_e P_f
\end{align}
with
\begin{align*}
	&\tilde{A}_0 = \tfrac{1}{2} K_{\alpha}K_{\beta}C_{\alpha\beta} \\ 
	&\tilde{A}_{ab} = A_{ab} + \tfrac{1}{2}\left(C_{\alpha \beta} H_{\beta ab}+C_{\beta \alpha} H_{\beta ab}- 2q_{\alpha ab}\right)K_{\alpha} \\
	&\tilde{A}_{abcd} = A_{abcd} + \tfrac{1}{2}\left(H_{\beta ab}C_{\beta \alpha} - 2q_{\alpha ab}\right)H_{\alpha cd} \, .
\end{align*}

Note that the tensors are no longer totally symmetric, since we have chosen a special direction (the wall-normal $\bs$). After some algebra, we can compare our general results to the many special cases considered in \cite{MRH2010}, and they match. The key advantage of our formalism is that there is a reasonably simple chain from the material coefficients $A, C, Q$ to the effective model for a given vacuum and wall orientation \eqref{eq:finalF}. The final analytic expressions for $\tilde A$ are incredibly complicated, especially when we allow for arbitrary walls. Hence, when writing code to solve the problem, we do not deal with them directly. Instead, we start with $A, C$ and $Q$ and implement the results of this section.

The initial potential $V^o$ is symmetric under all elements of the cubic group. The final potential depends on both the Polarisation vacuum value at both ends of the wall, $\boldsymbol{P}^{V_\infty}$ and $\boldsymbol{P}^{V_{-\infty}}$, and the wall-normal $\boldsymbol{s}$. The vectors define lines, and we can mark points on a cube where these lines intersect it. The symmetry of the problem is equal to the symmetry of the marked cube. For example, consider the wall along $\boldsymbol{s} = (0,0,1)$ joining $\boldsymbol{P}^V \propto (1,1,1)$ to $-\boldsymbol{P}^V$. The corresponding potential will be invariant under $\pi$ rotations about $(1,-1,0)$, a rotation by $\pi$ about $(0,0,1)$ combined with a reflection about the $z=0$ plane, and combinations of these. Overall, a $C_2 \times C_2$ subgroup of $O_h$.

Although somewhat obscured by the notation, the potential $V$ is a polynomial of order 6 and so domain walls in ferroelectrics are described by a multicomponent $\phi^6$ theory. This theory, originally proposed in \cite{lohe1979}, has been widely studied owing to its highly non-trivial ``fractal" dynamics between domain walls \cite{dorey2011, gani2014}. Multicomponent theories have been shown to exhibit even more complex dynamics \cite{alonso2020}. Unfortunately, the dynamics governing DWs in ferroelectrics are first order, so these studies are not directly applicable to this work.

\section{Static walls}

For the rest of the paper we will consider a specific ferroelectric, barium titanate. The material constants are given in Appendix A. The system has cubic symmetry and there are four phases with different global minima. These are:
\begin{itemize}
	\item $T<201K$, the rhombohedral phase, with $P^V = (p^r,p^r,p^r)$ plus permutations.
	\item $201K<T<282K$, the orthorhombic phase, with $P^V = (p^o,p^o,0)$ plus permutations.
	\item $282K<T<400K$, the tetragonal phase, with $P^V=(p^t,0,0)$ plus permutations.
	\item $400K < T$, the spherical phase, with $P^V = (0,0,0)$.
\end{itemize}
The permutations are any symmetry elements of the cubic group. So there are 8, 12 and 6 vacua for the rhomboedhral, orthorhombic and tetragonal phases. Throughout this work we choose $T=150, 250$ and $300$ K as representative tempatures of each nontrivial phase. The vacuum values $p^i$ can be calculated analytically using \eqref{eq:vaceq}, but the expressions are unpleasant. 

The allowed walls satisfy \eqref{eq:constcond}, which we consider to be an equation in terms of the wall-normal $\bs$ through
\begin{equation}
	Q_{abcd} = R(\bs)_{ae}R(\bs)_{bf}R(\bs)_{cg}R(\bs)_{dh}Q^o_{efgh} \, .
\end{equation}
The allowed combinations of $ P^{V_{-\infty}}$, $P^{V_{\infty}}$ and $\bs$ are displayed in Table \ref{tab:walls}. The wall names are given by the angle between the connected vacua. We see that \eqref{eq:constcond} is quite restrictive and only some walls are allowed.

\begin{table} 
	\begin{tabular}{l | c | c | c  | c }
		Name & $P^{V_{-\infty}}$ & $P^{V_{\infty}}$  & Wall ($\bs$) & Energy\\ \hline
				R71$^\circ$ & $(p,p,p)$ & $(-p,p,p)$  & $(1,0,0)$ &  0.509\\
		& & & $(0,1,1)$ &  0.117 \\
		R107$^\circ$ & $(p,p,p)$ & $(-p,-p,p)$  & $(1,1,0)$ & 0.536 \\
		& & & $(0,0,1)$ & 0.240 \\
		R180$^\circ$ & $(p,p,p)$ & $(-p,-p,-p)$ & All allowed & \\ \hline
		O60$^\circ$ & $(p,p,0)$ & $(p,0,p)$ &  $(0,-1,1)$  & 0.193  \\
		O90$^\circ$ & $(p,p,0)$ & $(p,-p,0)$  &$(1,0,0)$ & 0.232 \\
		& & & $(0,1,0)$ &  0.992 \\
		O120$^\circ$ & $(p,p,0)$ & $(0,-p,p)$  & $(1,0,1)$   & 0.435\\
		O180$^\circ$ & $(p,p,0)$ & $(-p,-p,0)$ &  All allowed &  \\ \hline 
T90$^\circ$ & $(p,0,0)$ & $(0,p,0)$  & $(1,1,0)$ &  0.405 \\
& & & $(1,-1,0)$ & 0.405 \\
T180$^\circ$ & $(p,0,0)$ & $(-p,0,0)$  & All allowed & \\ \hline
	\end{tabular}
	\caption{The allowed walls in a defect-free ferroelectric. We take the temperature to be 150, 250 and 300 $K$ for the rhombohedral, orthorhombic and tetragonal phases respectively. The energy is in dimensionless units; it can be converted to Joules through the factor $\sqrt{G_{11}^3(P^V)^4/A_{11}}$.} \label{tab:walls}
\end{table}

We now find minimal energy domain walls for a given pair of vacua and wall orientation. We first consider those vacua with a discrete set of allowed walls: the R71$^\circ$, R107$^\circ$, O60$^\circ$, O90$^\circ$, O120$^\circ$ and T90$^\circ$ walls. The first task is to find appropriate initial data. We expect the wall, which connects two minima, to pass through (or near) a saddle point or maxima of the potential. We can classify all such points and label them as $\bP^{V_A}$. Then we create initial data which looks like a wall linking $\bP^{V_{-\infty}}$ and $\bP^{V_{\infty}}$ through $\bP^{V_A}$, such as
\begin{equation} \label{eq:initial}
	 \tfrac{1}{2}(1-\tanh(s))\bP^{V_{-\infty}} + \tfrac{1}{2}(1+\tanh( s)) \bP^{V_{\infty}} + \sech(s)\bP^{V_A}
\end{equation}
 We then flow these initial data using the Arrested Newton Flow method, with fixed boundary data. The idea, originally proposed in \cite{battye1997}, is to replace the first order gradient flow with second order time dynamics
\begin{equation} \label{eq:gradflow}
	\ddot{P}_a = - \frac{\delta F}{\delta P_a}, \quad \bP(-\infty) = \bP^{V_{-\infty}}, \quad \bP(\infty) = \bP^{V_\infty}\, .
\end{equation}
The initial configuration will accelerate towards an energy minimum but, without interference, will not stop there. We interfere by checking the energy of the configuration at each time step and only proceed if the energy falls. If the energy increases, we set the field velocity to zero and restart the time evolution. Again the field will move towards a minima. Once we are close to a solution (the variation is below some critical tolerance) we stop the flow and are at a energy minimiser. The method is much faster than standard gradient flow, as the second order dynamics allow for much larger time steps.

We repeat this process for many initial configurations, including ones which pass through each possible $\bP^{V_A}$ and save the result with lowest energy. The energies of these configurations are written in Table \ref{tab:walls}. We have switched to dimensionless units by dividing $P$ by $|P^V|$, the $A$s by $A_{11}$ and the $G$s by $G_{11}$. The length and energy units are then given by $\sqrt{G_{11}/A_{11}}$ and $\sqrt{G_{11}^3(P^V)^4/A_{11}}$. Note that these units depend on temperature.

Certain wall directions, connecting the same vacua, are energetically preferred. For instance, O90$^\circ$ walls between $(p^o,p^o,0)$ and $(p^o,-p^o,0)$ in the $(1,0,0)$ direction are preferred over those in the $(0,1,0)$ direction. This confirms the result of \cite{Hlinka2006}, that head-to-tail walls are preferred over head-to-head and tail-to-tail walls. The reason is that the ``derivative" energy, $G_{sasb}\partial_sP_a\partial_bP_b$/2, of the head-to-tail wall is proportional to $G_{44}$, which is much smaller than of the head-to-head wall, which is proportional to $G_{11}$. 

\begin{figure}[h!]
	\begin{center}
		\includegraphics[width=\columnwidth]{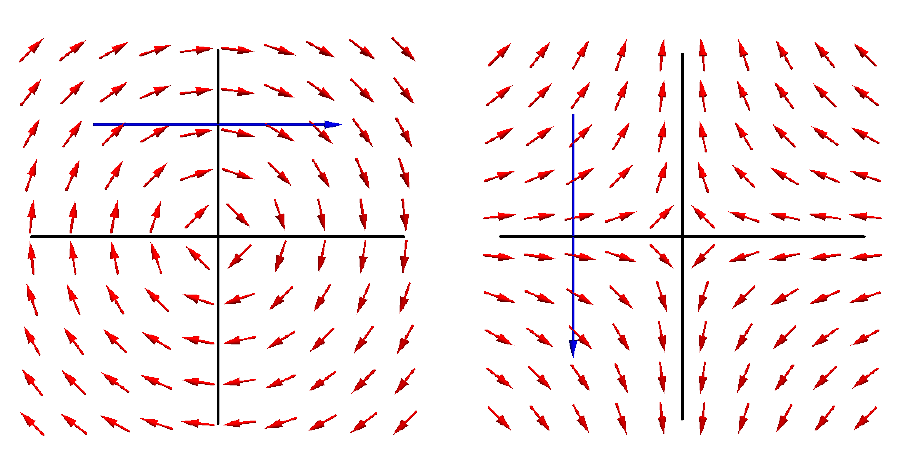}
		\caption{Two possible 2D structures. We can think of the vortex (left) as made from low energy head-to-tail walls (e.g. along the wall-normal highlighted in blue) connecting the regions separated by black lines. The antivortex (right) is made from high energy head-to-head and tail-to-tail walls. Our analysis of domain wall energies suggest that the vortex will have lower energy. }
		\label{fig:vantiv}
	\end{center}
\end{figure}

That certain wall directions have lower energy has implications for higher-dimensional objects. Consider the two 2D configurations in Figure \ref{fig:vantiv}. The first, a vortex, should be energetically preferred to the second, an anti-vortex, because it is constructed from walls with lower energy. This explains why the Ising Lines from \cite{stepkova2015} have a preferred (positive) chirality.

\section{Static $180^\circ$ walls}

\begin{figure}[h!]
	\begin{center}
		\includegraphics[width=\columnwidth]{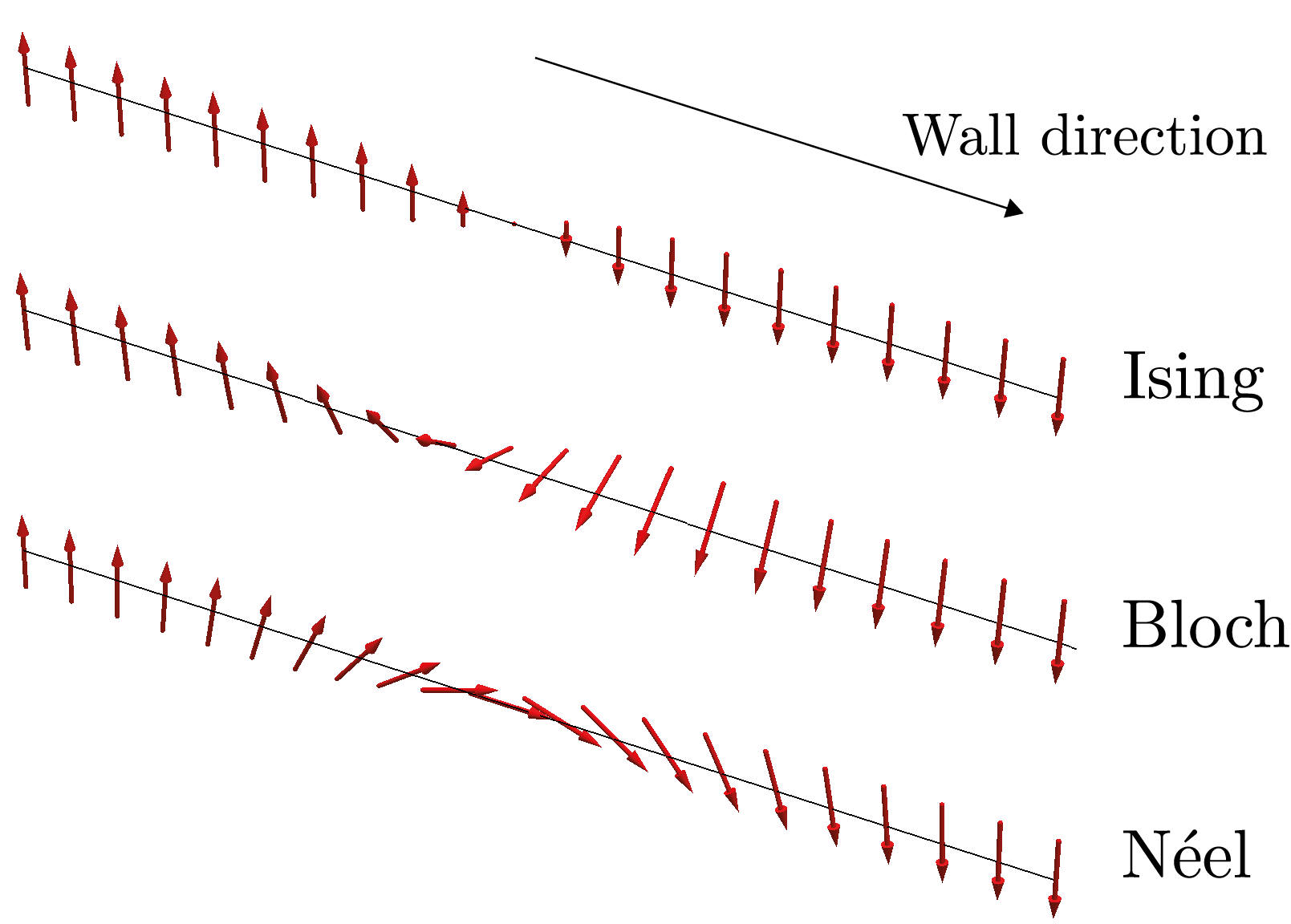}
		\caption{The three prototypical domain walls. }
		\label{fig:BIN}
	\end{center}
\end{figure}

The 180$^\circ$ walls are allowed to be in any orientation. For a given wall orientation, the most important question is: is the wall Ising, Bloch or Néel? In Ising walls, the Polarisation vector vanishes at its center. The Polarisation in Bloch walls varies perpendicularly to the wall-normal while Néel walls have non-zero Polarisation in the direction of the wall-normal. The three types are shown in Figure \ref{fig:BIN}. Pure Bloch and Néel walls are only allowed when the wall-normal $\boldsymbol{s}$ is orthogonal with $\boldsymbol{P}^V$. However, even when this is not the case, we can define the Ising, Néel and Bloch components of the Polarisation vector as
\begin{equation} \label{eq:Pibn}
	\bP(s) = P_I(s) \hat{\boldsymbol{P}}^V + P_N(s) \tilde{\boldsymbol{s}} + P_B(s) \hat{\boldsymbol{P}}^V \times \tilde{\boldsymbol{s}}   \, ,
\end{equation}
where $\tilde{\boldsymbol{s}}$ is the linear combination of $\boldsymbol{P}^V$ and $\boldsymbol{s}$ which is orthogonal to $\boldsymbol{P}^V$. In many cases, $\boldsymbol{P}^V$ and $\bs$ are automatically orthogonal. In this case, $\tilde{\bs}$ is simply equal to $\bs$. Given a generic wall the components $P_I, P_B$ and $P_N$ tell us which type of basic  wall it is closest to. The deconstruction \eqref{eq:Pibn} is only allowed if $\boldsymbol{P}^V$ and $\boldsymbol{s}$ are not equal. 

The rarest wall type is Bloch. Hence, we will search for a Bloch wall when varying the wall orientation. We can gain some insight by looking at the potential energy \eqref{eq:finalF} at the center of the wall, when $P_I = 0$. That is, we look at the energy
\begin{equation} \label{eq:VNB}
	V(P_N, P_B) = V\left( \boldsymbol{P} = P_N  \tilde{\boldsymbol{s}} + P_B   \hat{\boldsymbol{P}}^V \times \tilde{\boldsymbol{s}} \right) \, .
\end{equation}
This is a sextic polynomial in two variables whose coefficients depend non-trivially on the wall orientation (as explained in Section II). Hence we can classify its stationary points semi-analytically. A minimum at $(P_N, P_B) = (0,p_b)$ suggests that the system could support a Bloch wall while a minimum at $(P_N, P_B) = (p_n,0)$ suggests there could be a Néel wall. Of course, the derivative energy is also important, but this potential provides a theoretical starting point.

\begin{figure}[h!]
	\begin{center}
		\includegraphics[width=\columnwidth]{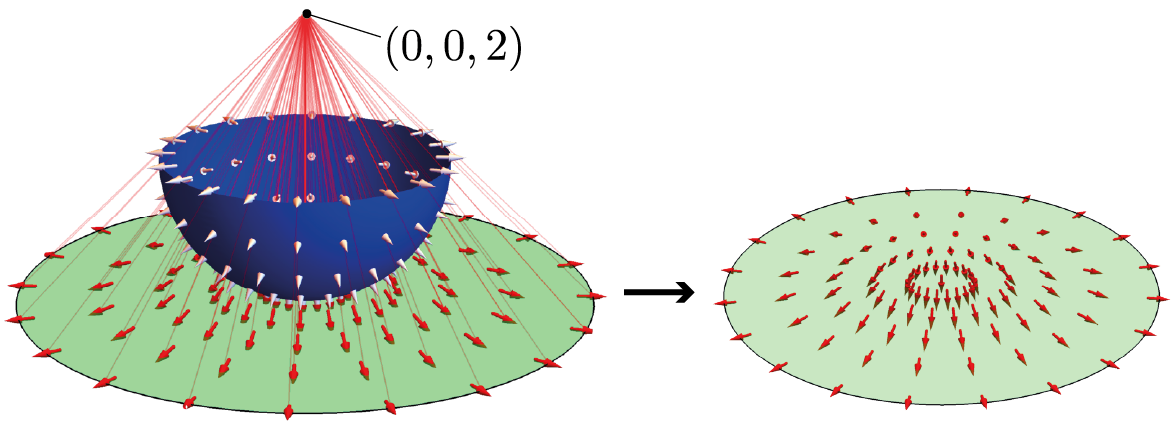}
		\caption{A visualisation of the projection map used in Figure \ref{fig:BvsI}. We take a sphere whose normal vector are the wall-normals $\bs$ (white arrows, left) and project these from $(0,0,2)$ to a disk on the plane $z=0$ (right).}
		\label{fig:proj}

\vspace*{0.5cm}

  \includegraphics[width=\columnwidth]{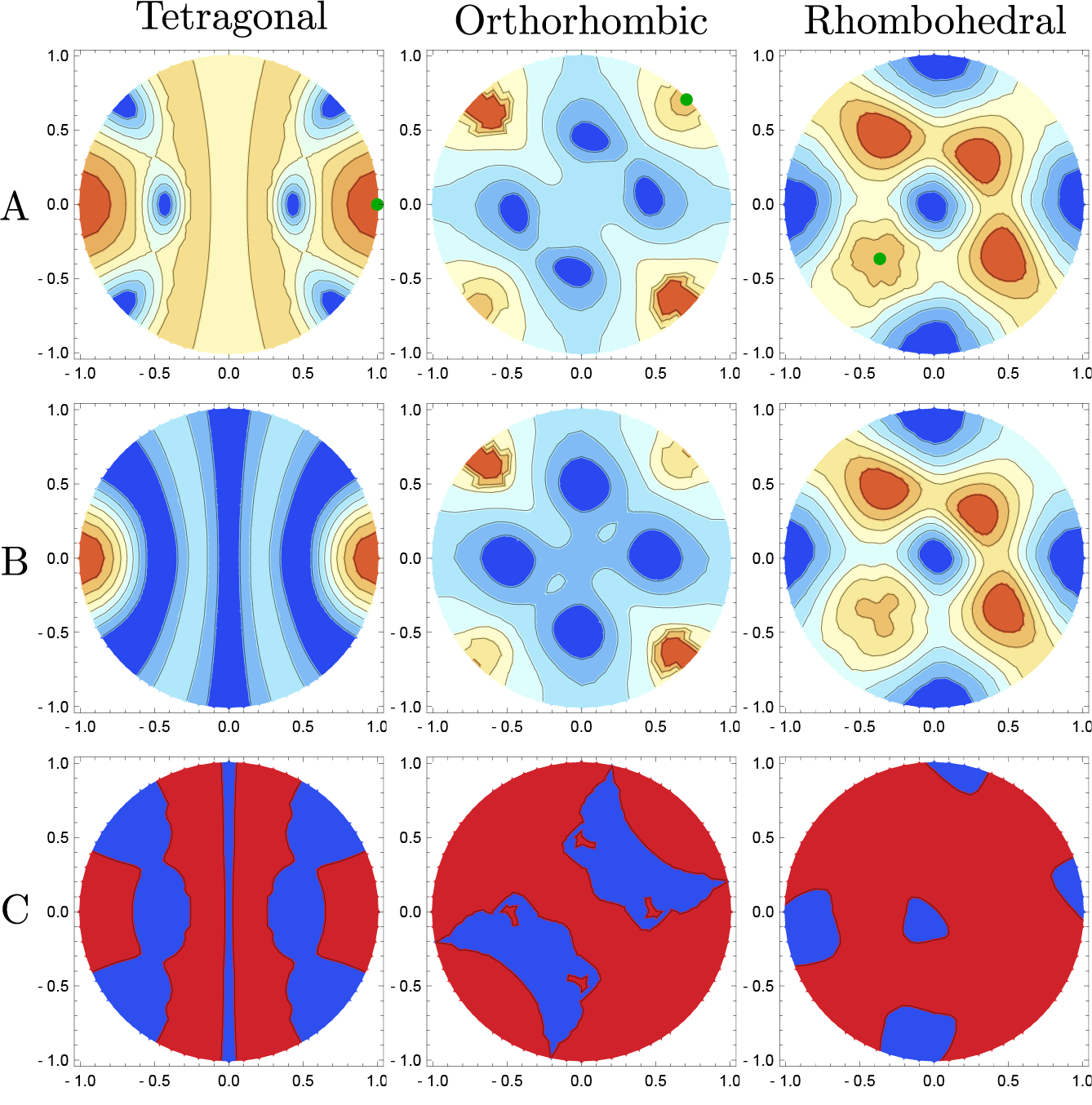}
		\caption{Plot of the wall-energy as a function of wall-normal $\bs$. Each point on each disk corresponds to a different unit wall-normal, with the mapping show in Figure \ref{fig:proj}. Red and blue represents large and small energies respectively. The minimal energy wall (row A), minimal energy wall subject to the Ising condition $\bP(0)= \bzero$ (row B) and difference between the energies (row C) are shown. The red regions contains wall-directions where the non-Ising wall has at least 10\% less energy. The tetragonal, orthorhombic and rhombohedral walls are between the vectors $(p^t,0,0)$ to $(-p^t,0,0)$, $(p^o,p^o,0)$ to $(-p^o,-p^o,0)$ and $(p^r,p^r,p^r)$ to $(-p^r,-p^r,-p^r)$, shown as a green dot in row A, at temperatures $300$ K, $250$ K and $150$ K respectively.  }
		\label{fig:BvsI}
	\end{center}
\end{figure}

We now search for wall orientations which support interesting walls and study them. To do so, let us change to coordinates which reflect the choice of wall orientation through the angles $\theta$ and $\phi$:
\begin{align}
\bs &= \left(\cos(\theta), \sin(\theta)\sin(\phi), \sin(\theta)\cos(\phi)\right) \\
\br &= \left(\sin(\theta), -\cos(\theta)\sin(\phi), -\cos(\theta)\cos(\phi)\right) \, .
\end{align}
 To begin, we find the overall energy minimiser for the given wall-orientation following the procedure described in the previous section for non-$180^\circ$ walls. We also find the minimal energy Ising wall by demanding that $\boldsymbol{P}(0) = \boldsymbol{0}$. We do this to check when non-Ising walls are favoured over Ising walls. These results are shown in Figure \ref{fig:BvsI}. We plot the energy of the energy-minimising wall in row A and the energy of the energy-minimising Ising wall in row B. In row C the blue region represents where the non-Ising wall has $10\%$ lower energy than the Ising wall. Here, the wall-orientation is parameterised using a Riemann half-sphere projected onto a disk in the plane. The projection is displayed graphically in Figure \ref{fig:proj}. The final disk is rescaled to have unit radius. In total, we found the energy minimising domain wall for 30,000 different wall orientations and initial conditions.

For a given $(\theta, \phi)$ pair, we can find the minima of the potential \eqref{eq:VNB}. If this analysis suggests there could be a Bloch or Néel wall, we study the example in more detail. This is how we have found the example discussed in the remainder of this section.

\begin{figure}[h!]
	\begin{center}
		\includegraphics[width=\columnwidth]{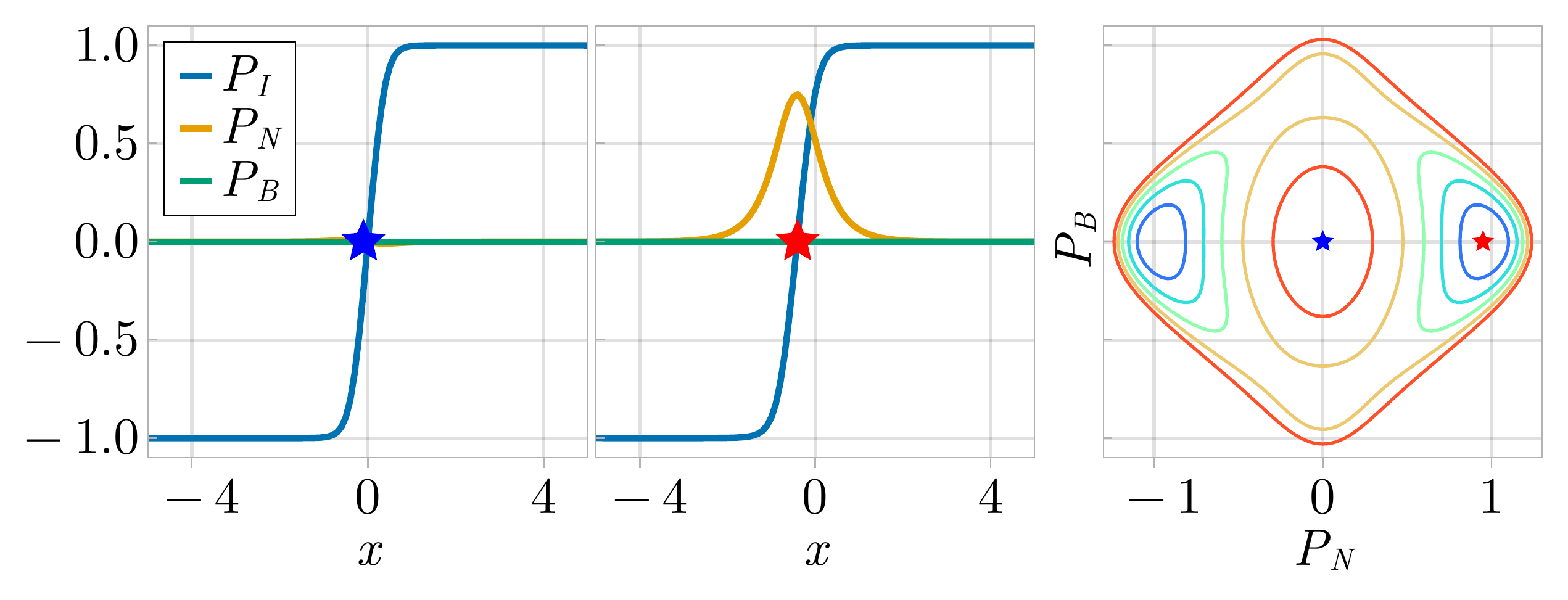}
		\caption{ The Ising wall (left), minimal energy Néel wall (middle) and a contour plot of the potential energy \eqref{eq:VNB} for $T=300$K, $\bP^V = (p^t,0,0)$ and $\boldsymbol{s} = (1/\sqrt{2},1/2,1/2)$. The center-point of each wall is marked with a star. The Ising, Bloch and Néel components of the wall, defined in \ref{eq:VNB} are plotted in blue, orange and green respectively. }
		\label{fig:tet}
	\end{center}
\end{figure}

In the tetragonal phase (left column of Figure \ref{fig:BvsI}) the wall connects vacua pointing towards cube faces: we take $(p^t,0,0)$ and $(-p^t,0,0)$. For Ising walls, the energy only depends on the angle between the wall and vacua, and the energy is minimised when the wall is perpendicular to the vacuum states. Non-Ising DWs are preferred in several orientations. An interesting example is when $\boldsymbol{s} = (1/\sqrt{2},1/2,1/2)$. The potential energy at the center of the wall \eqref{eq:VNB} for this orientation is plotted in Figure \ref{fig:tet} (middle), and suggests there could be a Bloch or Néel wall, since there are minima along these axes. It costs less potential energy to pass through a minima of the potential at the wall-center but it costs gradient energy to move the Polarisation away from $(P_N, P_B) = (0,0)$. In this case, the Néel wall has lowest energy: the cost of gradient energy to pass through the Néel point is lower than the potential energy saved. However, there does exist a higher energy, locally minimal, Ising wall. This example highlights the utility of the potential energy plots as they can predict which walls may have low energy and the importance of fitting the gradient coefficients $G$ correctly. If $G$ is larger, it would cost more energy to create the Néel wall and thus it could be disfavoured compared to the Ising wall.

\begin{figure}[h!]
	\begin{center}
		\includegraphics[width=\columnwidth]{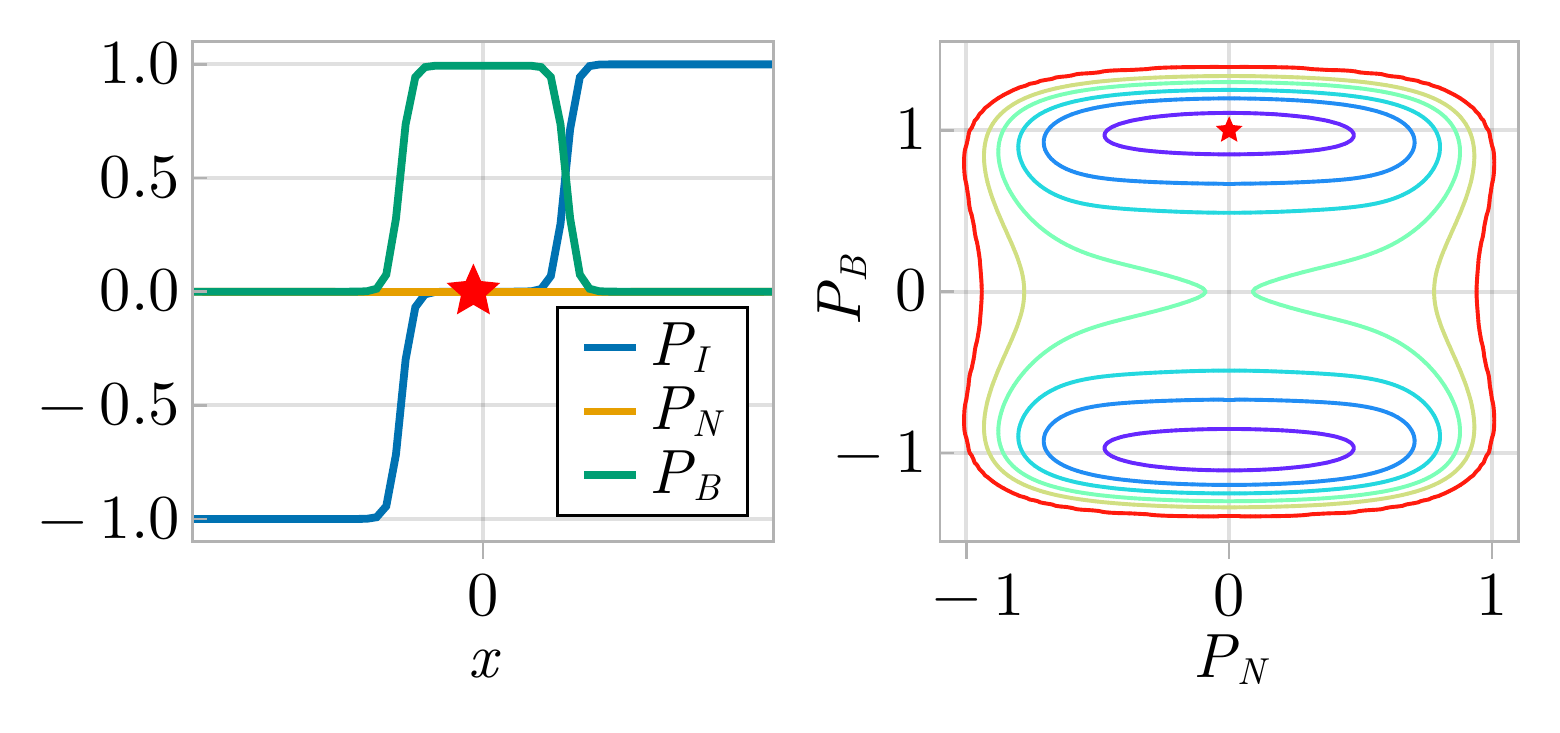}
		\caption{ The minimal energy $90^\circ-90^\circ$ wall for $T=250$K, $\bP^V = (p^o,p^o,0)$ and $\boldsymbol{s} = (0,0,1)$ (left) and a contour plot of the potential energy (right)  }
		\label{fig:ort}
	\end{center}
\end{figure}

In the orthorhombic phase, our DWs connect the vacua $(p^o,p^o,0)$ and $(-p^o,-p^o,0)$. By examining Figure \ref{fig:BIN}, we see that the lowest energy orientations are on the points of a diamond with, e.g. $\boldsymbol{s} = (0,1,1)/\sqrt{2}$. Usually the non-Ising walls are pairs of smaller degree walls. For example, when $\boldsymbol{s}=(0,0,1)$ the potential energy suggests that there is a possible Bloch wall. When constructed, the wall looks like two $90^\circ$ walls, one joining $(1,1,0)$ to $(1,-1,0)$ and another from $(1,-1,0)$ to $(-1,-1,0)$. The potential and solution can be seen in Figure \ref{fig:ort}. Later we will show that these $90^\circ$  walls repel one another. Similarly, the energy-minimising walls with, e.g., $\bs = (0,1,1)/\sqrt{2}$ are pairs of repelling $90^\circ$ Néel walls.

\begin{figure}[h!]
	\begin{center}
		\includegraphics[width=\columnwidth]{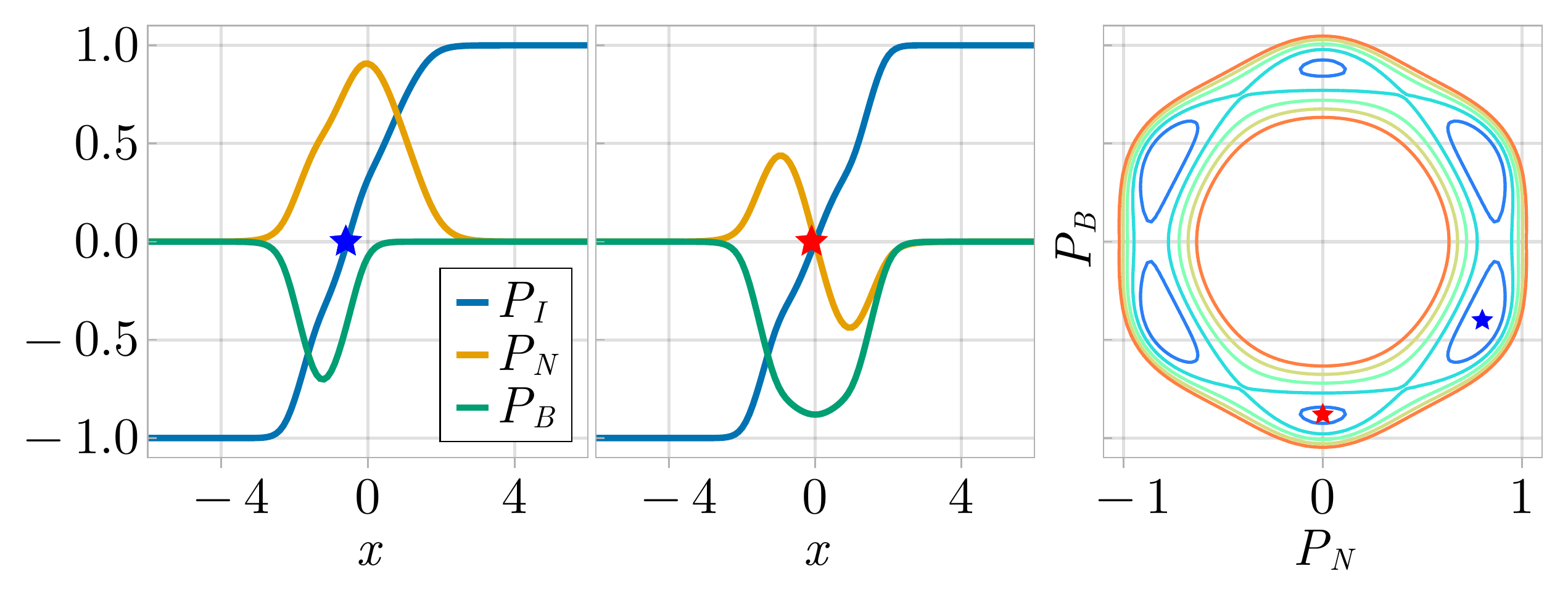}
		\caption{The minimal energy mixed wall (left) and higher energy Bloch wall (middle) in the rhombohedral phase at $T=150$ K and $\boldsymbol{s} = (1/\sqrt{2},-1/2-1/2)$. The potential energy for the wall-center (middle); the mixed wall passes close to the point $(P_N, P_B) \approx (0.8,-0.4)$ while the Bloch wall passes close to the point  $(P_N, P_B) \approx (0,-0.9)$. }
		\label{fig:rho}
	\end{center}
\end{figure}

Finally in the rhombohedral phase, the domain walls connect $(p^r,p^r,p^r)$ and $(-p^r,-p^r,-p^r)$. The energy minimising walls have cartesian normals such as  $(1,0,0)$. Non-Ising walls are preferred in some orientations, and some of these have been discovered previously \cite{taherinejad2012}. In this phase, we do find Bloch wall solutions, although they are not energy-minimising. One example is shown in Figure \ref{fig:rho}. The potential energy has minima in six places: two Bloch points and four mixed points which together form an approximate hexagon. The mixed points have lower energy than the Bloch points. We plot a solution which pass near the Bloch point (middle). Note that these are not pure Bloch walls, but the Bloch component is large compared to the Néel component. We also plot the solution which passes near the mixed point (left) and this has lower energy. It passes near vacua which lie on the corners of the cube.

Overall, we have found a variety of domain wall solutions. The minima of the potential energy \eqref{eq:VNB} can suggest where certain types of walls can form, but their existence and energy depend on various factors including the depth of the minima and the size of the derivative energy. There are non-Ising walls preferred in every phase of the system.

\section{Asymptotics and Interactions} \label{sec:asyandints}

We will now calculate the asymptotic form of a domain wall, and use this to understand their interaction. This will allow us study wall-wall pairs and show that, in general, Ising-Ising walls are unstable to collapse but other pairs may not be.

Consider the tail of a wall, where
\begin{equation}
	\bP = \bP^V + \bp
\end{equation}
and $p_a$ is small. The Euler-Lagrange equations for the tail are
\begin{equation}
-G_{sasb}\partial_s^2 p_b + \frac{\partial^2 V}{\partial P_a \partial P_b }\bigg \lvert_{P^V} p_b = 0\, ,
\end{equation}
which has solution
\begin{equation}
	\bp =  \sum_i \bmu_i e^{-\lambda_i s} \,
\end{equation}
where $\lambda_i^2$ and $\bmu_i$ are the eigenvalues and eigenvectors of
\begin{equation} \label{eq:gddv}
	G^{-1}_{sasc} \frac{\partial^2 V}{\partial P_c \partial P_b }\bigg \lvert_{P^V} \, .
\end{equation}
The leading behaviour is described by the eigenvector with the smallest eigenvalue. 

Now consider two walls, one going from $\bP^{V_{-\infty}}$ to  $\bP^{V_0}$ and another from $\bP^{V_0}$ to $\bP^{V_{\infty}}$. Suppose the walls have positions $\mp X$ with $|X| >>1$, and denote them $\bP^{\mp X}$. Near the central vacuum, the walls take the form $\bP^{\mp \bX}(s) = \bP^{V_0} + \bp^{\mp \bX}(s)$. We can approximate the combined wall as
\begin{align}
	\bP(s) &= \bP^{-X}(s) + \bP^X(s) - \bP^{V_0} \\
	 &\approx \begin{cases} \bP^{-X}(s) + \bp^X(s) - \bP^{V_0} \quad \text{for $s<0$} \\
		\bp^{-X}(s) + \bP^X(s) - \bP^{V_0} \quad \text{for $s>0$} \end{cases}
\end{align}
Using this approximation, we can evaluate the energy of the configuration as a series in $\bp$. The calculation is done in detail in \cite{halcrow2022stable}, and the result only depends on the tails:
\begin{equation} \label{eq:Eint}
	E(p)^\text{int} = G_{sasb}\left(p_b^X \partial_s p_a^{-X}  - p_b^{-X} \partial_s p_a^X \right) \, .
\end{equation}
This simple expressions tells us if the walls attract or repel based solely on their tails. We'll now examine three cases.

First, consider two `short' walls which combine to make long wall, like the two $90^\circ$ walls seen in  Figure \ref{fig:ort}. Here, $\bs = (0,0,1)$ and we can choose $\br = (1,1,0)/\sqrt{2}$ so that the vacuum is given by $\bP = p^r\br$. The smallest-eigenvalue  eigenvector of \eqref{eq:gddv} at $s=0$ is $\be_s$. So the tails of the $90^\circ$ walls are given by
\begin{equation}
	\bp^{-X} = a \be_s e^{-\lambda(s+X)}\, , \, \bp^X = -a \be_s e^{\lambda(s-X)} \, ,
\end{equation}
where we've chosen signs so that $a >0$. Then
\begin{equation}
	E^\text{int} =  2 a^2 G_{ssss} \lambda^2 e^{-2\lambda X} \, .
\end{equation}
The combined gradient term and eigenvalues are positive and so, regardless of the absolute value of the decay $a$, the interaction energy is positive. The walls can decrease the energy by increasing $X$, revealing that the walls repel. We use fixed boundary data, explaining the result seen in Figure \ref{fig:ort}. 

Let us consider two $180^\circ$ Ising walls: the first connecting a vacuum $\bP^{V}$ to its negative $-\bP^V$ and the second $-\bP^V$ back to $\bP^V$. For simplicity, we'll consider an explicit example. Choose coordinates $(\bs,\br,\bt)$ so that the wall-normal is $\bs$ and the connected vacua are in the $\br$ direction so that $\bP^{V_\infty} = (0,P^V,0)$.  The wall tails at $s=0$ are given by
\begin{align}
\bp^{-X} = b \be_r \exp(-\lambda (s+X))\, , \quad \bp^{X} =  b\be_r \exp(\lambda (s-X)) 
\end{align}
where
\begin{equation}
	\lambda^2 = G_{srsr}^{-1} \frac{\partial^2 V}{\partial P_r \partial P_r }\bigg \lvert_{-\bP^V} \, .
\end{equation}
Note that, since $-\bP^V$ is a vacuum, the Hessian is positive. Using \eqref{eq:Eint} we find
\begin{equation}
	E^\text{int} = -2b^2\lambda G_{srsr} \exp\left(-2\lambda X\right) \, .
\end{equation}
The eigenvalue and $G_{srsr}$ are positive and so the energy is negative. The walls can decrease the energy by decreasing $X$ and so the walls attract and will, eventually, annihilate. This result is well known in kink models: kinks and antikinks always attract in one-component theories. The Ising wall ansatz reduces our multicomponent theory to a single component one.

\subsection{Domain wall clusters}

Until now, we have studied walls which either attract or repel each other. But since the theory is multicomponent, there could be pairs which attract at long range but repel at short range. There could then be stable wall clusters. We used the asymptotic analysis to search for wall-orientations which support long range attraction. We then evolved the system to see if there were stable solutions.

One example with stable domain wall clusters is the configuration joining $\boldsymbol{P}^V = p^o(1,1,0)$ to $-\boldsymbol{P}^V$ and back again along the wall with $\boldsymbol{s} = (1,1,1)/\sqrt{3}$. This has dominant eigenvector $(0,0,1)$ and the wall tails in this channel create an attractive force at long range. The second most dominant eigenmode creates a repulsion. The balance between the attraction and repulsion in different eigenchannels generates the conditions for stable configurations. Note that the final configuration traces a loop around the equator in target space. For the walls to collapse, this loop must contract. But this costs significant energy because both the point $\bP = \bzero$ and the lines $P_z \neq 0$ have high energy. This energy barrier stops the loop from becoming a point, and hence the system is stable.

\begin{figure}[h!]
	\begin{center}
		\includegraphics[width=01.0\columnwidth]{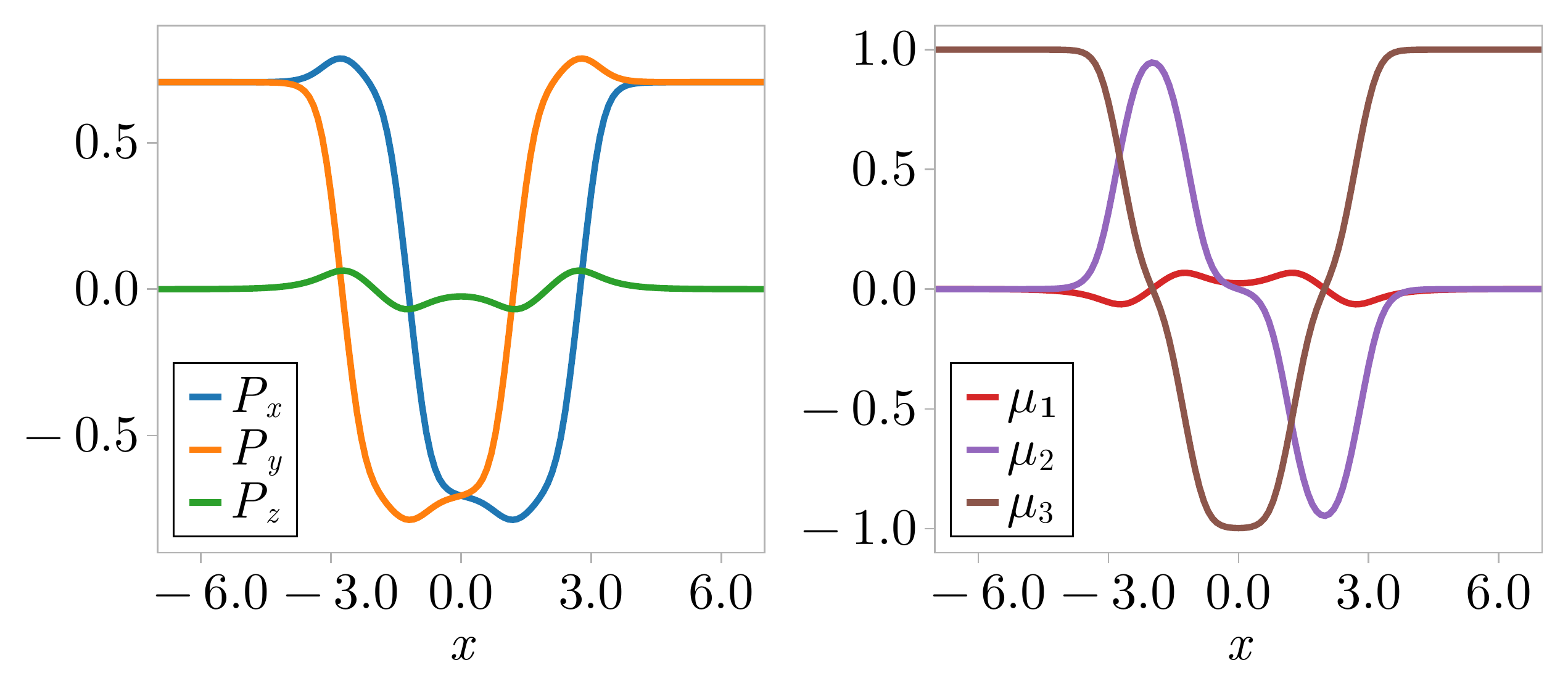}
		\caption{A configuration with long range attraction but short range repulsion, allowing for a stable local minimum. We plot the field in $xyz$-components and in the asymptotic eigenbasis. There is attraction in the first and second eigen-components and repulsion in the third.}
		\label{fig:wwmin}
	\end{center}
\end{figure}

\begin{figure}[h!]
	\begin{center}
		\includegraphics[width=1.0\columnwidth]{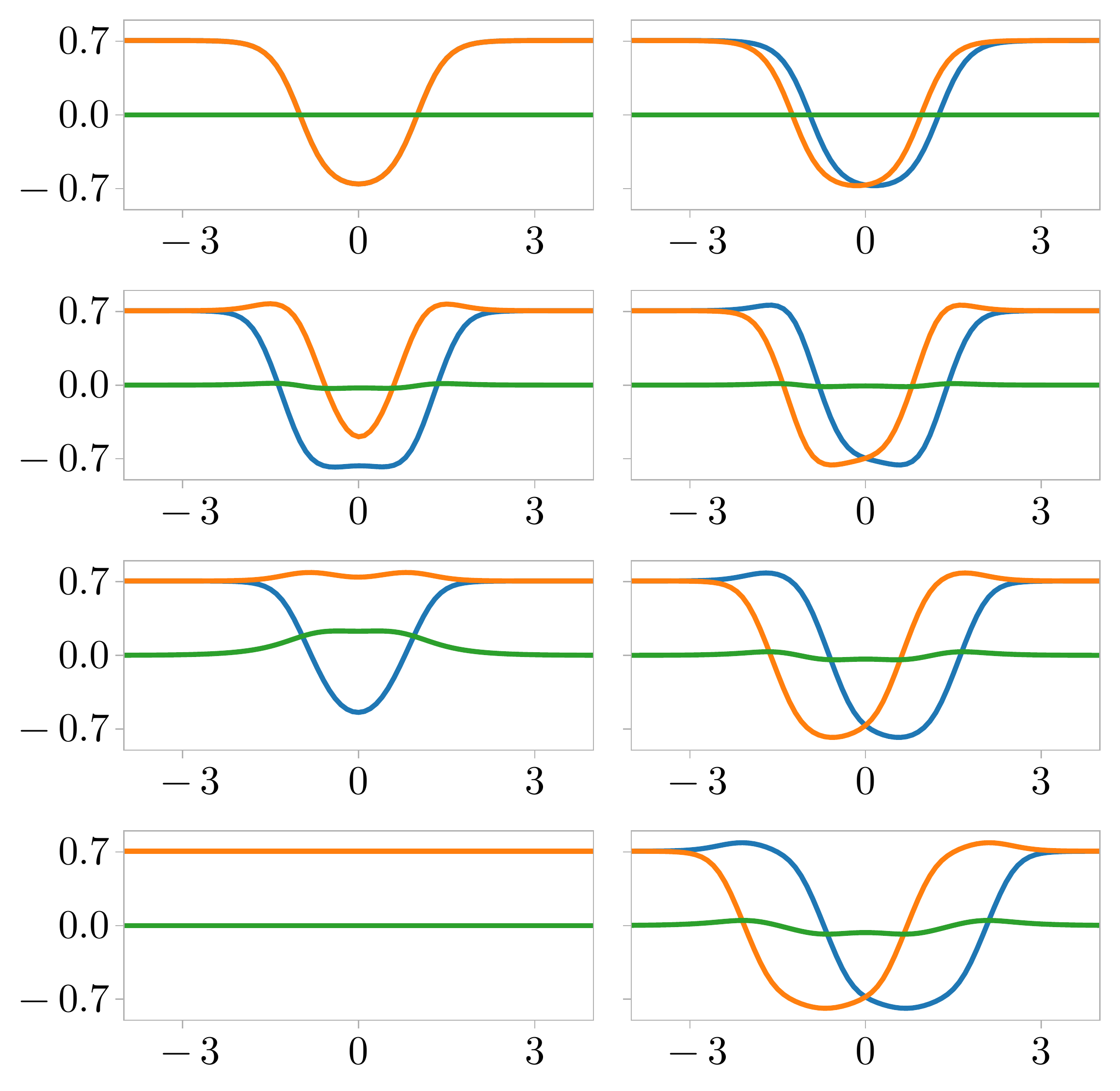}
		\caption{Gradient flow applied to initial data representing Ising-Ising and non-Ising pairs in the orthorhombic phase at $T=200$K, for the wall-normal $\bs = (1,1,1)/\sqrt{3}$ and boundary Polarisation $\bP^{\infty} = (p^o,p^o,0)$. We plot the $xyz$-components of the Polarisation vector, coloured in the same way as in Figure \ref{fig:wwmin}. Time is read from top to bottom. The Ising-Ising pair attract and annhilate into the vacuum while the non-Ising walls fall into a stable minima.}
		\label{fig:IIvsNN}
	\end{center}
\end{figure}

Both unstable and stable wall-wall pairs can be constructed for the same wall-orientation and phase. We construct initial data based on two Ising wall and two non-Ising walls in the top of Figure \ref{fig:IIvsNN}. We create the Ising walls by demanding that the field vanishes at two points. We then remove this constraint and flow the field. The time evolution is read from top to bottom. As predicted by the asymptotic analysis, the Ising walls annhilate while the non-Ising walls stabilise. This mechanism could explain why certain domain wall pairs are more stable than others: it could be because they are non-Ising.

\begin{figure}[h!]
	\begin{center}
		\includegraphics[width=1.0\columnwidth]{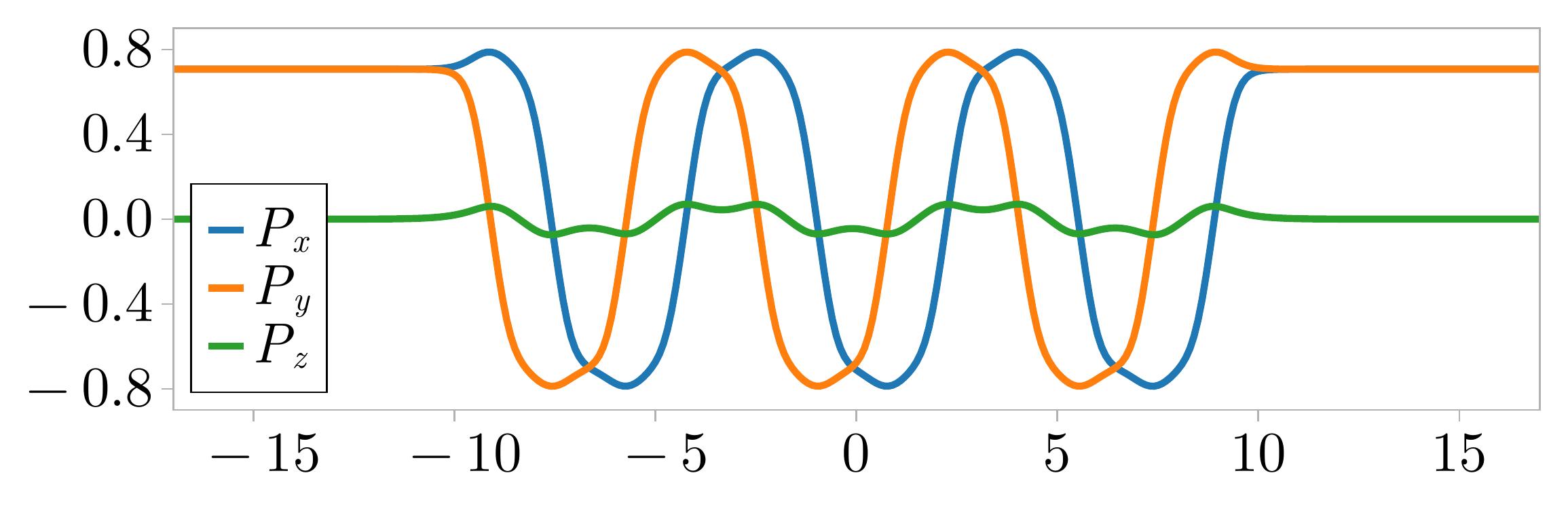}
		\caption{A stable cluster formed of 6 domain walls.}
		\label{fig:6chain}
	\end{center}
\end{figure}

We can also look for longer clusters of domain walls but adjusting the initial conditions for the gradient flow. For example, a stable 6-cluster is displayed in Figure \ref{fig:6chain}.

\begin{figure}
	\begin{center}
        \includegraphics[width=1.0\columnwidth]{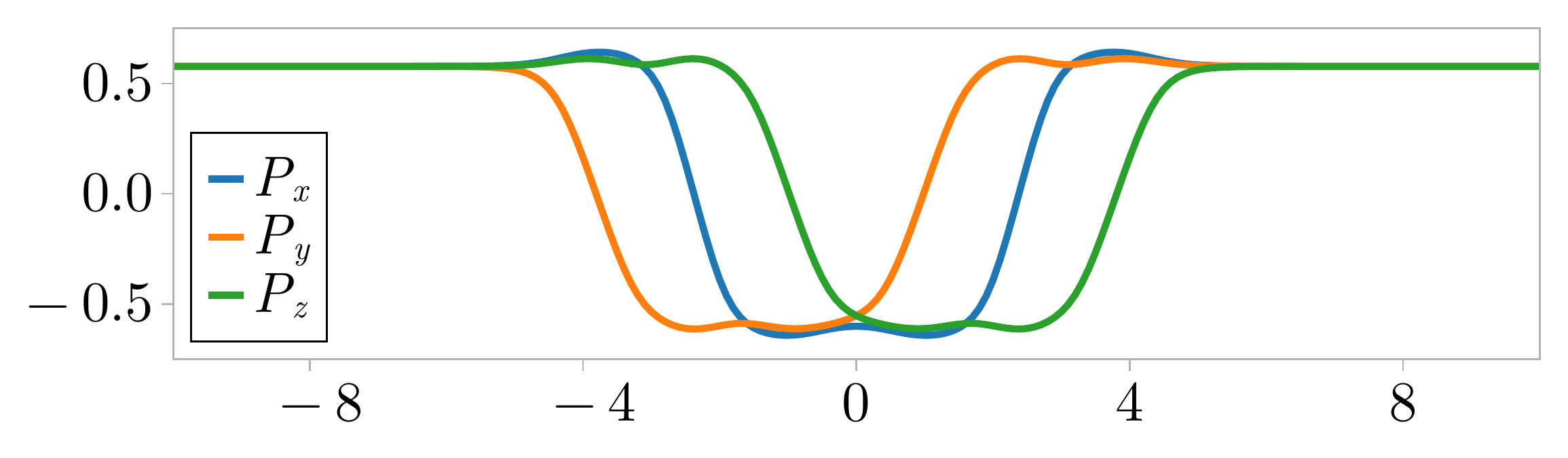}
		\caption{A stable wall-wall pair in the rhombohedral phase at $T=150$ K. Here, $\boldsymbol{s} = (1,-1,-1)/\sqrt{3}$. The order of the components is vital for ensuring the stability of this configuration. }
		\label{fig:rhopair}
	\end{center}
\end{figure}

Generally, it is rare to find wall-orientations which support long range attraction for the wall pairs. However, there are many examples, especially in the rhombohedral phase, where the different eigenmodes give separately attractive and repulsive interactions. In these situations it is possible that there are stable minima, such as the configuration in Figure \ref{fig:rhopair}, due to nonlinear effects. The long-range attraction is required to prove analytically that these are truly minima. The other walls we can numerically show that they are stable by calculating the Hessian around the configuration. We have done this for the field shown in Figure \ref{fig:rhopair}.

\section{Linear and nonlinear stability}

Given a solution to the static equations of motion $\bP^0(s)$, we can study its linear stability through its normal modes, $e_a(s)$. These satisfy
\begin{equation}
\hat{H}[\bP^0(s)]_{ab}e_b = \omega^2 e_a
\end{equation}
where $\hat{H}$ is given by
\begin{equation}
\hat{H}_{ab} = -G_{sasb}\partial_s \partial_s + \frac{\partial^2 V}{\partial P_a \partial P_b}\bigg\rvert_{\bP^0(s)}
\end{equation}
This is a Schr\"odinger type equation and methods to find its solutions are well known. We do so using a gradient flow method developed in \cite{barnes1997} and applied to higher dimensional system in \cite{halcrow2018}. Briefly, we take a random initial perturbation $e$ and evolve it using
\begin{equation}
	\partial_\tau e = -\hat{H} e \, .
\end{equation}
This has solution
\begin{equation}
	e(s) = \sum_n \exp(- \omega_n t) e_n(s)
\end{equation}
where $e_n$ and $\omega_n$ are eigenfunctions and eigenvalues. After a long time, the lowest energy mode will dominate the flow while all other modes are exponentially suppressed. We save the lowest mode and repeat the process while projecting out our saved mode. The second lowest frequency mode then dominates the solution at large times and we save this. And so on.

Let us consider an example: the stable wall-wall pair found in the previous section, seen in Figure \ref{fig:wwmin}. Denote the solution $P^{W}(s)$. It has a $\mathbb{Z}_2$ reflection symmetry
\begin{equation*}
(P^{W}_x(s), P^{W}_y(s), P^{W}_z(s) ) = (P^{W}_y(-s), P^{W}_x(-s), P^{W}_z(-s) )
\end{equation*}
and so we can label the modes depend on whether they transform trivially under this transformation, or pick up a sign. We find the four lowest frequency normal modes with frequencies and signs  $\omega^P = 0.0^-, 0.2^+, 1.34^+$ and $1.35^-$. All modes are positive, confirming that the wall-pair is a local minimum. Physically, the first mode corresponds to translations and hence costs no energy to excite. In the second, the walls oscillate around their positions: moving towards and away from each other. The final two modes are breathing modes: the walls increase and decrease in size, either in- or out-of-phase.

\begin{figure}
	\begin{center}
		\includegraphics[width=1.0\columnwidth]{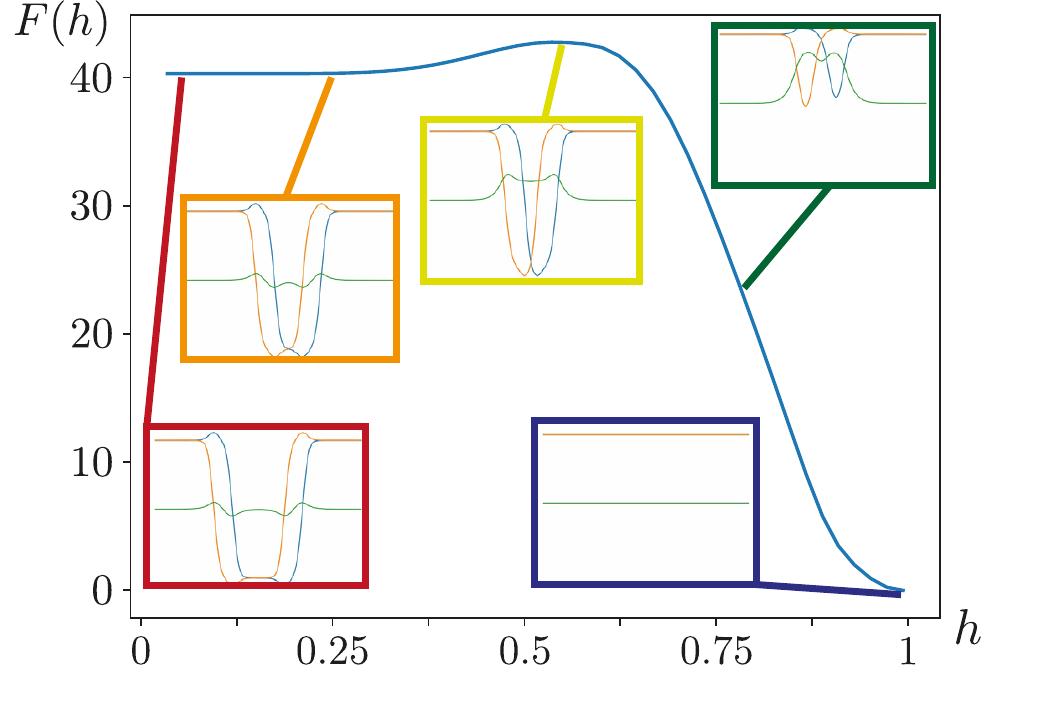}
		\caption{The energy $F(P(h))$ of our minimised band of configurations as a function of the distance along the string $h$, joining the minimum $P^{W}(s)$ to the vacuum $\boldsymbol{P}^V$ for $T=200$K. Several configurations along the band are plotted, using the same colouring and axes are used in Figure \ref{fig:wwmin}. The band passes through a local minimum ($h\approx0.25$), saddle point ($h\approx0.6$).}
		\label{fig:saddle}
	\end{center}
\end{figure}

We know that there is a configuration with lower energy than the wall-pair: the vacuum $P(s) = P^V$. Hence, although the configuration is linearly stable, it cannot be stable under large perturbations. We can use the Simplified String Method to understand the nonlinear stability. This method was originally proposed for chemical reactions \cite{weinan2007}, but has also been used 
 study stability of topological defects in condensed matter \cite{benfenati2020}. We construct a ``string" of $n$ configurations which interpolate between the wall-pair and the vacuum, parameterised by $h_i = \frac{i-1}{n-1} \in [0,1]$:
\begin{equation}
P(h_i,s) =	(1-h_i)P^{W}(s) + h_i P_V \, .
\end{equation}
We then minimise the energy of the band
\begin{equation}
	\sum_{i=1}^n F\left[  P(h_i) \right] \, ,
\end{equation}
using a gradient flow, with the end points fixed as
\begin{equation}
	P(0,s) = P^{W}(s) \quad P(1,s) = P^V \, .
\end{equation}
In the simplified string method we demand that, during the flow, the points are evenly spaced on the string. In detail, we define a measure of distance
\begin{equation}
	d\left(P(h_i) , P(h_{j})\right)^2 = \int \left( P(h_i) - P(h_{j}) \right)^2 \, ds \, .
\end{equation}
Then the $k^{\text{th}}$ configuration is a distance 
\begin{equation}
	L_k = \sum_{i=1}^{k-1} 	d\left(P(h_i) , P(h_{i+1})\right) \, .
\end{equation}
along the string, and the total length is given by $L_n$. After we have flowed for some time, we recalculate the total length of the string and linearly re-sample the configurations by demanding that the configurations are evenly spaced, at positions $L^\text{new}_n = L_nh_i$. If $L_{k-1} < L^\text{new}_n < L_{k}$, then the re-sampled configuration at that point is given by
\begin{equation}
\frac{L_k - \tilde{L}}{L_k - L_{k-1}} P(h_{k-1},s) + \frac{\tilde{L} - L_{k-1}}{L_k - L_{k-1}}P(h_{k},s) \, .
\end{equation}
The process generates a low energy path in configuration space joining two minima which, by Morse theory, is guaranteed to pass through a saddle point. The energy of the saddle point gives the energy barrier to collapse. We re-apply the simplified string method with end configurations $P^W(s)$ and two widely separated walls. The energy $E(h_i)$ and some configurations $P(h_i,s)$ from these simulations are plotted in in Figure \ref{fig:saddle}.

In the path from $P^{W}$ to the vacuum, the configuration must either pass through $\bP = \bzero$ or contain a non-zero $P_z$. Both cost significant energy and, in this case, the lower energy option is to increase $P_z$.  On target space, the initial state is a circle while the end state is a point. So, during the process the circle has contracted to a point: the topology of the configuration has changed. The energy barrier exists because changing this topology costs energy (due to the high energy cost away from the $P_x^2 + P_y^2 = |P^V|^2$). Although it is difficult to see on the graph, there is a long-range attraction and the energy has a minima at $h\approx 0.25$.

\section{Creation of walls using external fields}

We have seen that non-Ising walls can exist and that there are stable non-Ising wall-wall pairs. We will now try to generate these interesting configurations starting from a simpler intial field and applying an external electric field. To model this, the term
\begin{equation}  \label{eq:Eext}
-\boldsymbol{E}(\boldsymbol{x},t) \cdot \bP \, ,
\end{equation}
is added to the free energy \eqref{eq:finalF}. The electric field encourages the Polarisation vector to point in its direction and so we can use it to construct any desired configuration. We will turn on a specific electric field for a short burst, then turn it off and continue to flow the fields. The process is considered a success if final state is our desired one.

We begin by taking an Ising wall and applying an electric field to try and change it into a non-Ising wall. We'll consider a case previously discussed: the tetragonal wall with orientation $\bs = (1/\sqrt{2},1/2,1/2)$ and $T=300$K first seen in Figure \ref{fig:tet}. In the Cartesian basis, the wall connects the vacua $(p^t,0,0)$ and $(-p^t,0,0)$. There is a stable Néel wall, with centre-value approximately equal to $1/\sqrt{2}(0,p^t,p^t)$. Hence our initial state is an Ising wall, and we apply the external field
\begin{equation}
    E_{ext}(s,t) = \begin{cases} (0,E,E) \quad &-s_0<s<s_0, t<t_0 \\ 0 \quad & \text{otherwise} \end{cases} .
\end{equation}
We take $t_0 = 5$ and $s_0 = 2.6$nm. The wall changes from an Ising to a Néel wall provided that $E>0.32$MV/m.

Next we consider the stable wall-wall pair $P^W$ seen in Figure \ref{fig:wwmin}. This exists in the orthorhombic phase at $T=250$K and wall $\bs = (1,1,1)/\sqrt{3}$. We'll take the initial state to be the constant vacuum. The final state passes near four vacua, from $(1,1,0)$ to $(1,-1,0)$ to $(-1,-1,0)$ to $(-1,1,0)$ (in units of $p^t$) and back to the start. We need a more complicated external field to generate the more complicated configuration. We take
\begin{equation}
    E_{ext}(s,t) = \begin{cases} (-E,0,0) \quad &-s_0<s<0, t<t_0 \\ 
    (E,0,0) \quad &0<s<s_0, t<t_0 \\
    0 \quad & \text{otherwise} \end{cases} .
\end{equation}
with $s_0 = 6$nm. The stable $P^W$ configuration is generated and stabilises provided $E>14 $MV/m. This is an order of magnitude larger than the field required to change the nature of a wall since we are creating walls from the vacuum.

\section{Summary}

In summary, we have calculated the energy-minimising domain walls for all possible wall orientations in barium titanate. 
In every phase, we found orientations which support non-Ising walls. Most of these were superpositions of shorter walls.
By studying the asymptotic form of the walls, we investigated the stability of multi-wall configurations. 
Most interestingly, we found the effect of domain wall clustering. 
Their stability can be explained by the existence of an effective topology. 
We developed tools to calculate their linear and nonlinear stability: allowing the study of their collapse. 
Finally, we found the required external electric field to create the non-Ising configurations.

The methods developed here can be applied to topological defects in higher dimensions with minimal adjustment. 

The stability of domain walls clusters, as well as the fact that they can be created on demand by the protocol that we outline, suggests that the walls are not only the objects of academic interest, but also can be used in memory applications.

Our formalism for domain walls in arbitrary orientations applies to all ferroelectric Ginzburg-Landau-Devonshire models, simply with different material constants. Hence it is simple to repeat the analysis for other materials such as PZT, lithium-niobate and lithium-tantalate.

\begin{acknowledgments}
We thank Katia Gallo, Anton Talkachov, Albert Samoilenka, Mats Barkman and Sahal Kaushik for useful discussions. CH is supported by the Carl Trygger Foundation through the grant CTS 20:25. This work is supported by the Swedish Research Council Grants 2016-06122 and 2022-04763
and by the Knut and Alice Wallenberg
Foundation through the Wallenberg Center for Quantum Technology (WACQT).
\end{acknowledgments}

\appendix*

\section{Material constants}

In this appendix we write down the material constants used in the text, and how their commonly used form relates to our tensors.

We take the potential material constants $\alpha$ derived in \cite{buessem1966}, except for $\alpha_{123}$ which comes from \cite{bell1984}. The elastic constants $C, Q$ (and hence $q$) are the same as those used in \cite{Hlinka2006}. There are slightly improved constants available in \cite{zgonik1994} but we felt it was more important to have our work directly comparable to similar theoretical studies. There is more uncertainly in the value of the the derivative energy tensor $G$. We use the parameters proposed in \cite{Hlinka2006} (note that our $\alpha_{11}$ and $\alpha_{12}$ is equal to their $\alpha^{(e)}_{11}$ and $\alpha^{(e)}_{12}$). All the parameters can be found in Table \ref{tab:constants}.

We bundle the parameters into tensors as follows
\begin{align*}
    &A_{ij} = \delta_{ij}\alpha_1 \\
    &A_{ijkl} = \left(\alpha_{11} - \tfrac{1}{2}\alpha_{12} \right) \delta_{ijkl} + \tfrac{\alpha_{12}}{6} \left(\delta_{ij}\delta_{kl} + \delta_{ik}\delta_{jl} + \delta_{il}\delta_{jk} \right) \\
    &A_{ijklmn} = \left( \alpha_{111}-\alpha_{112} + \tfrac{1}{3}\alpha_{123}\right) \delta_{ijklmn} \\
     &\qquad + \tfrac{1}{6}a_{123}\delta_{ij}\delta_{kl}\delta_{mn} \\
     &\qquad + \left(\tfrac{1}{15}a_{112} - \tfrac{1}{30}a_{123} \right)\left( \delta_{ij}\delta_{klmn} + \text{14 perms} \right) \, .
\end{align*}
The tensors $C$ and $G$ have the same decomposition. We give it for $C$,
\begin{align*}
    C_{ijkl} &= \left(C_{11} - C_{12} - 2C_{44}\right)\delta_{ijkl} + C_{12}\delta_{ij}\delta_{kl} \\
    &+ C_{44}\left(\delta_{ik}\delta_{jl} + \delta_{il}\delta_{jk} \right) \, ,
\end{align*}
while $q$ and $Q$ are similar but slightly modified,
\begin{align*}
    q_{ijkl} &= \left(q_{11} - q_{12} - q_{44}\right)\delta_{ijkl} + q_{12}\delta_{ij}\delta_{kl} \\
    &+ \tfrac{1}{2}q_{44}\left(\delta_{ik}\delta_{jl} + \delta_{il}\delta_{jk} \right) \, .
\end{align*}

\begin{table}[b]
	\begin{tabular}{l | c | l }
 Const. & Value & Units \\ \hline
		$\alpha_1$ & $3.34 (T-381)$ & $10^5$ JmC$^{-2}$\\
  $\alpha_{11}$ & $4.69(T-393) - 202$ & $10^6$ Jm$^5$C$^{-4}$ \\
  $\alpha_{12}$ & 3.23 & $10^8$ Jm$^5$C$^{-4}$ \\
  $\alpha_{111}$ & $-5.52(T-393) + 2760$ & $10^6$ Jm$^9$C$^{-6}$ \\
  $\alpha_{123}$ & 4.91 & $10^9$ Jm$^9$C$^{-6}$  \\
  $\alpha_{112}$ & 4.47 & $10^9$ Jm$^9$C$^{-6}$  \\ \hline
  $C_{11}$ & 2.75 & $10^{11}$ Jm$^{-3}$ \\
  $C_{12}$ & 1.79 & $10^{11}$ Jm$^{-3}$ \\
  $C_{44}$ & 5.43 & $10^{10}$ Jm$^{-3}$ \\ \hline
  $q_{11}$ & 1.42 & $10^{10}$ JmC$^{-2}$ \\
  $q_{12}$ & -7.4 & $10^{8}$ JmC$^{-2}$ \\
  $q_{44}$ & 1.57 & $10^{9}$ JmC$^{-2}$ \\ \hline
  $G_{11}$ & 51 & $10^{-11}$ Jm$^3$ C$^{-2}$ \\
  $G_{12}$ & -2 & $10^{-11}$ Jm$^3$ C$^{-2}$ \\
  $G_{44}$ & 2 & $10^{-11}$ Jm$^3$ C$^{-2}$ \\ 
	\end{tabular} 
	\caption{The material constants used in this paper.} \label{tab:constants}
\end{table}

\bibliographystyle{ieeetr}
\bibliography{DWs.bib}{}

\end{document}